\documentclass[aps, prstab, reprint, showpacs, amsmath, amssymb]{revtex4-1}

\usepackage{amsmath,amscd}
\usepackage{bm}% bold math
\usepackage[english]{babel}
\usepackage{graphicx}% Include figure files
\usepackage{soul}
\usepackage[usenames]{color}
\renewcommand{\Re}{\mathop{\mathrm{Re}}\nolimits}

\begin{document}
\title{ Diffraction at the Open End of Dielectric-Lined Circular Waveguide }

\author{Sergey N. Galyamin}
\email{s.galyamin@spbu.ru}
%\author{Andrey V. Tyukhtin}
\affiliation{Saint Petersburg State University, 7/9 Universitetskaya nab., St. Petersburg, 199034 Russia}
\author{Viktor V. Vorobev}
\affiliation{Technical University of Munich, Germany, Department of Informatics}

\date{\today}

\begin{abstract}
A rigorous approach for solving canonical circular open-ended dielectric-lined waveguide diffraction problems is presented.
This is continuation of our recent paper~\cite{GVT2021} where a simpler case of uniform dielectric filling has been considered.
Here we deal with the case of an open-ended circular waveguide with layered dielectric filling which is closer to potential applications. 
The presented method uses the solution of corresponding Wiener-Hopf-Fock equation and leads to an infinite linear system for reflection coefficients (S-parameters) of the waveguide, the latter can be efficiently solved numerically using the reducing technique.
As a specific example directly applicable to beam-driven radiation sources based on dielectric-lined capillaries, diffraction of a slow TM symmetrical mode at the open end of the described waveguide is considered.
A series of such modes forms the wakefield (Cherenkov radiation field) generated by a charged particle bunch during its passage along the vacuum channel axis.
Calculated S-parameters were compared with those obtained from COMSOL simulation and an excellent agreement was shown.
This method is expected to be very convenient for analytical investigation of various electromagnetic interactions of Terahertz (THz) waves (both free and guided) and charged particle bunches with slow-wave structures prospective in context of modern beam-driven THz emitters, THz accererators and THz-based bunch manipulation and bunch diagnostic systems.
\end{abstract}

%\begin{IEEEkeywords}
%Diffraction radiation, open-ended waveguide, Wiener-Hopf technique
%\end{IEEEkeywords}

%\IEEEpeerreviewmaketitle

\maketitle
%\tableofcontents

\section{Introduction}

Modern applications of dielectric-lined waveguides, including both open-ended wavegiudes and resonators, are tightly connected with Cherenkov effect.
Cherenkov radiation (CR) has been initially discovered with fast electrons traversing dielectric medium and emitting radiation in the visible region of electromagnetic spectrum~\cite{Ch37}. 
Through decades, CR has been succesfully used for a variety of applications in high-energy physics~\cite{Zrb}.
Today considerable advances have been reached in implementation of CR effect for dielectric wakefield acceleration~\cite{OShea16} where CR in the form of a wakefield with up to GeV per meter magnitude and Terahertz (THz) frequencies can be generated by high-quality relativistic electron bunches passing through dielectric-lined waveguide structures (capillaries).
Segmented dielectric-lined waveguides were also offered to manipulate the longitudinal phase space of the bunch~\cite{Mayet2020}.  

%On the other hand, it has been far understood that radiation of the same nature occurs during the uniform movement of any localized source with the speed exceeding the phase velocity of electromagnetic waves in the given medium within the given frequency range~\cite{Askaryan1962}.
%Again, Cherenkov-type radiation produced by short moving pulses of polarization generated in nonlinear crystals via optical rectification of laser pulses~\cite{Bass1962, Auston84} is considered nowadays as a most advanced and versatile mechanism for producing wideband THz radiation~\cite{BakunovBodrov2020} especially in the tilted-pulse-front scheme~\cite{Hebling2020, BodrovBakunov2019}.

In recent years, contemporary beam technologies has became tightly interlaced with modern THz technologies.
The latter are actively developed due to unique properties of THz radiation having large amount of prospective application connected with precise manipulating and probing the state of the matter~\cite{Hebling2020}.  
Moreover, these technologies penetrate to beam physics: strong THz fields allow realization of THz driven electron guns~\cite{Kartner2015}, performing THz bunch compression, streaking~\cite{AntipovXiang2020, Nanni2020} and wakefield acceleration within THz driven dielectric-lined waveguide structures~\cite{Nanni2015, Pacey2020}. 
Inversely, beam technologies contribute to THz ones: dielectric capillaries similar to those used for the THz bunch manipulation can be in turn utilized for development of high-power narrow-band THz sources~\cite{WangAntipov2018, GTAB14}.

It is worth noting that almost all the mentioned cases involve interaction of both THz waves (free or guided) and charged particle bunches with an open end of certain waveguide structure loaded with dielectric, most frequently a circular capillary~\cite{OShea16, AntipovXiang2020}.
For further development of the discussed prospective topics a rigorous approach allowing analytical investigation of both radiation from open-ended capillaries and their excitation by external source (bunch or electromagnetic pulse) would be very useful.
In our recent paper~\cite{GVT2021}, we have presented an efficient rigorous method for solving circular open-ended waveguide diffraction problems and illustrated this method using the case of uniform dielectric filling of the waveguide.
Here we deal with more realistic case of a layered filling (vacuum channel and dielectric layer) and internal excitation by single waveguide mode.
Moreover, though the presented technique can be rigorously extended to the beam-driven case (similar to how it has been done for ``embedded'' structures~\cite{GTVGA19}), it can be applied approximately to the CR in the form of a narrow-band wakefield generated behind the driver bunch.

%This is especially important, for example, for the case of high-order CR modes generation resulting in THz wakefields in mm-sized capillaries~\cite{GTAB14} since corresponding numerical simulations can be over-complicated.
%However, up to now such an approach was absent despite the fact that general diffraction theory for open-ended wavegiude discontinuity was intensively developed.
%
%We present an efficient rigorous method for solving circular open-ended dielectric-lined waveguide diffraction problems.
%Here we deal with the typical case of a two layers (vacuum channel ans dielectric layer) and internal excitation by single waveguide mode.
%Moreover, though the presented technique can be rigorously extended to the beam-driven case (similar to how it has been done for ``embedded'' structures~\cite{GTVGA19}), it can be applied approximately to the CR in the form of a narrow-band wakefield generated behind the driver bunch.
%It is also notable that though the discussed approach is valid for the orthogonal waveguide end only whereas an inclined cut (Vlasov antenna) is often more convenient for practice, the obtained analytical results can be both extremely useful for improvement of corresponding approximate methods~\cite{GTAB14, IGTT14} and serve as reference points for numerical simulations.
% 
\section{Problem formulation and general solution}

We consider an open-ended semi-infinite cylindrical waveguide with radius $ a $ lined with a dielectric of thickness $ a - b $ so that the region $ z < 0 $, $ b < \rho < a $ is filled with dielectric $ \varepsilon > 1 $ (Fig.~\ref{fig:geom}).
Both the region outside the waveguide ($ z > 0 $ and $ z < 0 $, $ \rho > a $) and the channel inside the waveguide ($ z < 0 $, $ \rho < b $) are filled with vacuum.
Waveguide walls are supposed to have an ideal electric conductivity.

The electromagnetic (EM) problem is solved in the frequency domain so that Fourier integral decomposition is used. 
For example, for $H_{ \varphi }$ component (cylindrical frame $\rho ,\varphi ,z$ is used) we have:
\begin{equation}
H_{ \varphi } = \int\nolimits_{-\infty}^{+\infty } H_{\omega \varphi } e^{ -i \omega t }\,d\omega.
\end{equation}
The problem is formulated for $ H_{\omega \varphi } $ while other nonzero field components can be derived as follows: 
\begin{align}
\label{eq:ErhoEz}
E_{ \omega \rho  } &= \frac{ 1 } {i k_0 \varepsilon } \frac{ \partial H_{\omega \varphi } }{\partial z }, \\
E_{ \omega z  } &= \frac{ i } { k_0 \varepsilon }
\left( \frac{ H_{\omega \varphi } }{ \rho } + \frac{ \partial H_{\omega \varphi } }{\partial \rho } \right). 
\end{align}
In particular, we have 
$ E_{ \omega z } = 0 $
for 
$ \rho = a $, 
$ z < 0 $.

We suppose that single symmetrical $T{{M}_{0l}}$ waveguide mode incidents the orthogonal open end:
%
%%%%%%%%%%%%%%%%%%%%%%%%%%%%%%
\begin{figure}
\centering
\includegraphics[width=0.4\textwidth]{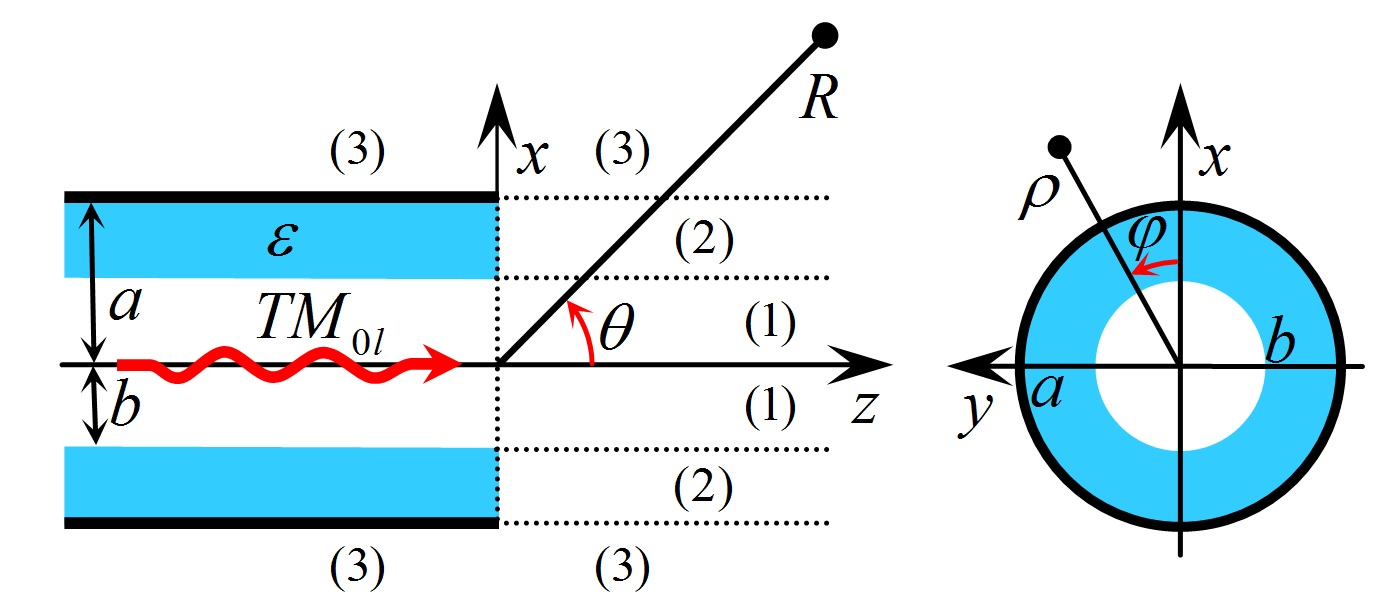}
\caption{\label{fig:geom} Geometry of the problem and main notations.}
\end{figure}
%%%%%%%%%%%%%%%%%%%%%%%%%%%%%%
%
%
\begin{equation}
\label{eq:Hphii}
H_{\omega \varphi }^{(i)}
=
M^{(i)} e^{ i k_{ z l } z }
\left\{
\begin{aligned}
&\left. J_1 ( \rho \sigma_l ) \right/ \sigma_l \quad \text{for  } \rho < b, \\
& \left[ J_1( \rho s_l ) Y_0( a s_l ) - Y_1( \rho s_l ) J_0( a s_l ) \right] \times \\
& \times \frac{ J_1 ( b \sigma_l ) }{ \sigma_l \psi_0( s_l ) } 
\quad \text{for  } b < \rho < a, 
\end{aligned}
\right.
\end{equation} 
where
$ M^{(i)} $ is an arbitrary amplitude constant for the incident mode,
$ J_{ \nu } $ and $ Y_{ \nu } $ are Bessel and Neumann functions of $\nu$-th order, correspondingly.
Transverse wave numbers $ \sigma_m $ and $ s_m $ are determined by the following dispersion equation
\begin{equation}
\label{eq:disp}
\varepsilon \sigma_m J_0( b \sigma_m ) \psi_0( s_m )
=
s_m J_1( b \sigma_m ) \psi_1( s_m ),
\end{equation}
where
\begin{align}
\label{eq:Abel}
\psi_0( s_m ) &= J_1( b s_m ) Y_0( a s_m ) - J_0( a s_m ) Y_1( b s_m ), \\
\psi_1( s_m ) &= J_0( b s_m ) Y_0( a s_m ) - J_0( a s_m ) Y_0( b s_m ) 
\end{align}
are so-called Abel functions.
Longitudinal wave number $ k_z $ is connected with $ \sigma_m $ and $ s_m $ as follows:
\begin{equation}
\label{eq:longwn}
k_{ z m } = \sqrt{ k_0^2 - \sigma_m^2 } = \sqrt{ k_0^2 \varepsilon - s_m^2 },
\quad
\mathrm{Im}\sqrt{\phantom{1}} > 0,
\end{equation} 
$ k_0 = \omega / c + i \delta$
(%
$ \delta \to 0 $,
which is equivalent to infinitely small dissipation an all areas), 
$ c $ is the light speed in vacuum.
From~\eqref{eq:longwn} one can express $ \sigma_m $ through $ s_m $ and obtain the dispersion relation~\eqref{eq:disp} with respect to a single variable $ s_m $.
Note that $ \sigma_0 = 0 $ is the solution of the dispersion equation~\eqref{eq:disp} if the following condition for the frequency holds:
\begin{equation}
\label{eq:dispsig0}
2 \varepsilon \psi_0( s_0 )
=
b s_0 \psi_1( s_0 ),
\end{equation}
and the corresponding waveguide mode propagates with the speed of light $ c $ because $ k_{ z 0 } = k_0 $.

%Connection between~\eqref{eq:Hphii} and CR wakefield generated by a charged particle bunch moving through the considered waveguide will be discussed below.

The reflected field in the area inside the waveguide ($ z < 0 $, $ \rho < a $) is decomposed into a series of waveguide modes propagating in the opposite direction:
\begin{equation}
\label{eq:r_fields}
\begin{aligned}
&H_{\omega \varphi }^{(r)}
=
\sum\limits_{ m = 1 }^{ \infty }
M_m e^{ - i k_{ z m } z } \times \\
&\times
\left\{
\begin{aligned}
&\left. J_1 ( \rho \sigma_m )\right/ \sigma_m \quad \text{for  } \rho < b, \\
& \left[ J_1( \rho s_m ) Y_0( a s_m ) - Y_1( \rho s_m ) J_0( a s_m ) \right] \times \\
& \times \frac{ J_1 ( b \sigma_m ) }{ \sigma_m \psi_0( s_m ) } 
\quad \text{for  } b < \rho < a, 
\end{aligned}
\right.
\end{aligned}
\end{equation}
where $\{ M_m \}$ are unknown ``reflection coefficients'' that should be determined.
The area outside the waveguide is divided into three subareas ``1'', ``2'' and ``3'' (see Fig.~\ref{fig:geom}), where the field is described by Helmholtz equation:
\begin{equation}
\label{eq:Helmholtz}
\left[
\frac{ \partial^{ 2 } }{ \partial z^2 }+\frac{ \partial^2 }{ \partial \rho^2 }+
\frac{ 1 }{ \rho } \frac{ \partial }{\partial \rho } +
\left( k_0^2 - \frac{ 1 }{ \rho^2 } \right)
\right]
H_{\omega \varphi}^{(1,2,3)}=0.
\end{equation}
We introduce functions ${{\Psi }_{\pm }}(\rho ,\alpha )$ (hereafter subscripts $\pm $ mean that function is holomorphic and free of poles and zeros in areas $\operatorname{Im}\alpha >-\delta$ and $\operatorname{Im}\alpha <\delta$, correspondingly):
\begin{equation}
\label{eq:plustrans}
\Psi _{+}^{(1,2,3)}(\rho ,\alpha )= {{(2\pi )}^{-1}}\int_{0}^{\infty }{dzH_{\omega \varphi}^{(1,2,3)}(\rho ,z){{e}^{i\alpha z}}},
\end{equation}
\begin{equation}
\label{eq:minustrans}
\Psi _{-}^{(3)}(\rho ,\alpha )={{(2\pi )}^{-1}}\int_{-\infty }^{0}{dzH_{\omega \varphi}^{(3)}(\rho ,z){{e}^{i\alpha z}}},
\end{equation}
and similar transforms of
$ E_{\omega z}^{(1, 2, 3 )} $,
for example,
\begin{equation}
\label{eq:Phi}
\Phi_{+}^{(1,2,3)}(\rho ,\alpha )={{(2\pi )}^{-1}}\int_{0 }^{\infty}{dz \frac{ k_0 }{ i } E_{\omega z}^{(1,2,3)}(\rho ,z){{e}^{i\alpha z}}}.
\end{equation}
From~\eqref{eq:Phi} and \eqref{eq:ErhoEz} we have the following relation between $\Phi$ and $\Psi$:
\begin{equation}
\label{eq:Psi2Phi}
\Phi_+^{ ( 1, 2, 3 ) }( \rho, \alpha )
=
\frac{ \Psi_+^{ ( 1, 2, 3 ) }( \rho, \alpha ) }{ \rho } + \frac{ \partial \Psi_+^{ ( 1, 2, 3 ) }( \rho, \alpha ) }{\partial \rho },
\end{equation}
and the same relation between $ \Phi_-^{ ( 3 ) }( \rho, \alpha ) $ and $ \Psi_-^{ ( 3 )( \rho, \alpha ) } $. 

From~\eqref{eq:Helmholtz} we obtain
\begin{equation}
\label{eq:Psi}
\left( \frac{{{\partial }^{2}}}{\partial {{\rho }^{2}}} {+} \frac{1}{\rho }\frac{\partial }{\partial \rho } {+} {{\kappa }^{2}} {-} \frac{1}{\rho^2} \right)
\left\{ \begin{matrix}
   \Psi _{+}^{(1,2)}  \\
   \Psi _{-}^{(3)}{+}\Psi _{+}^{(3)}  \\
\end{matrix}
\right\}{=}
\left\{
\begin{matrix}
   {{F}^{(1,2)}}  \\
   0  \\
\end{matrix} \right\},
\end{equation}
\begin{equation}
2 \pi F^{ ( 1, 2 ) } = 
{ \left. {\partial H_{\omega \varphi}^{(1,2)}}/{\partial z} \right| }_{z=+0} - {\left. i\alpha H_{\omega \varphi}^{(1,2)} \right|}_{ z=+0 },
\end{equation}
where $\kappa =\sqrt{k_{0}^{2}-{{\alpha }^{2}}}$, $\operatorname{Im}\kappa >0$.
Equation~\eqref{eq:Psi} is obtained as follows.
In subareas ``1'' and ``2'', we apply the integral operator
\begin{equation} 
\frac{ 1 }{ 2 \pi } \int\nolimits_0^{+\infty} \cdot {e}^{i\alpha z} \, d z
\end{equation}
to Eq.~\eqref{eq:Helmholtz}%
, use integration by parts and suppose that both 
$ H_{ \omega \varphi} $ 
and 
$ \partial H_{ \omega \varphi} / \partial z $ 
vanishes for 
$ z \to +\infty $.
In the issue we obtain the equations in the upper row of Eq.~\eqref{eq:Psi}.
Similarly, in the subarea ``3'', we apply the integral operator
\begin{equation} 
\frac{ 1 }{ 2 \pi } \int\limits_{-\infty}^{+\infty} \cdot {e}^{i\alpha z} \, d z =
\frac{ 1 }{ 2 \pi } \left( \int\limits_{-\infty}^{ 0 } + \int\limits_{ 0 }^{+\infty} \right) \cdot {e}^{i\alpha z} \, d z
\end{equation}
to Eq.~\eqref{eq:Helmholtz} 
and suppose additionally that 
$ H_{ \omega \varphi} $ 
and 
$ \partial H_{ \omega \varphi} / \partial z $ 
vanishes for 
$ z \to -\infty $,
in the issue we obtain the equation in the lower row of Eq.~\eqref{eq:Psi}.

Functions ${{F}^{(1,2)}}$ are determined using continuity of
$ E_{ \omega \rho } $ and $ H_{ \omega \varphi} $ at $ z = 0 $, $\rho < a $, 
therefore
\begin{equation}
\label{eq:BC}
\left. H_{ \omega \varphi} \right|_{z=+0} = \left. H_{ \omega \varphi } \right|_{z=-0},
\,\,
\left. \frac{ \partial H_{ \omega \varphi} }{ \partial z }\right|_{ z = +0 } = \left. i k_0 E_{ \omega \rho } \right|_{ z = -0 }, 
\end{equation}
while the right-hand sides of Eq.~\eqref{eq:BC}
can be calculated via the mode decomposition for reflected fields, see Eqs.~\eqref{eq:r_fields} 
and~\eqref{eq:ErhoEz}.
After transformations we obtain:
\begin{equation}
\label{eq:F1calc}
\begin{aligned}
2\pi{{F}^{(1)}} &= 
i \left[
\vphantom{\sum\nolimits_{m=1}^{\infty }} 
\left( k_{ z l } - \alpha \right)
M^{(i)} \left. J_1 ( \rho \sigma_l ) \right/ \sigma_l \right. \\
&\left.
-\sum\nolimits_{ m = 1 }^{ \infty } 
\left( k_{ z m } + \alpha \right)
M_m \left. J_1 ( \rho \sigma_m ) \right/ \sigma_m
\right], 
\end{aligned}
\end{equation}
\begin{equation}
\label{eq:F2calc}
\begin{aligned}
2 \pi F^{ ( 2 ) } 
&= 
i \left[
M^{(i)}
\left( \frac{ k_{ z l } }{ \varepsilon } - \alpha \right)
\frac{ J_1 ( b \sigma_l ) }{ \sigma_l \psi_0( s_l ) } \right. \times \\
& \times
\left[ J_1( \rho s_l ) Y_0( a s_l ) - Y_1( \rho s_l ) J_0( a s_l ) \right] - \\
&-\sum\nolimits_{m=1}^{\infty } M_m \left( \frac{ k_{zm} }{\varepsilon} + \alpha \right)
\frac{ J_1 ( b \sigma_m ) }{ \sigma_m \psi_0( s_m ) } \times \\
& \left. 
\times
\left[ J_1( \rho s_m ) Y_0( a s_m ) - Y_1( \rho s_m ) J_0( a s_m ) \right] 
\vphantom{ \left( \frac{ k_{ z l } }{ \varepsilon } - \alpha \right) }
\right]. 
\end{aligned}
\end{equation}
General solution of Eq.~\eqref{eq:Psi} has the form
\begin{align}
\label{eq:gen_sol_1}
\Psi_+^{ ( 1 ) }( \rho, \alpha ) 
&=
C^{(1)} J_1 ( \rho \kappa ) + \Psi_p^{ ( 1 ) }( \rho, \alpha ), \\
\label{eq:gen_sol_2}
\Psi_+^{ ( 2 ) }( \rho, \alpha ) &= C^{ ( 2 ) }_1 J_1 ( \rho \kappa ) + \nonumber \\
&+ C^{ ( 2 ) }_2 J_1( \rho \kappa ) + \Psi_p^{ ( 2 ) }( \rho, \alpha ), \\
\label{eq:gen_sol_3}
\Psi_-^{ ( 3 ) }( \rho, \alpha ) &+ \Psi_+^{ ( 3 ) }( \rho, \alpha ) = C^{ ( 3 ) } H_1^{ ( 1 ) }( \rho \kappa ),
\end{align}
where 
$H^{ ( 1 ) }_{ \nu }$
is a Hankel funcion of the first kind of $\nu$-th order,
$ C^{ ( 1, 3 ) } $ and $ C^{ ( 2 ) }_{ 1, 2 } $ are unknown coefficients, Eq.~\eqref{eq:gen_sol_1} 
contains only finite for 
$\rho \to 0 $ 
solution of the corresponding homogeneous equation, while Eq.~\eqref{eq:gen_sol_3} contains only the outgoing wave
$ \sim \exp( i \kappa \rho) $ for $\rho \to +\infty $.

Particular solutions of the inhomogeneous equations $ \Psi_p^{ ( 1, 2 ) } $ have the form
%%
%\begin{equation}
%\label{eq:Psi_p_guess}
%\Psi _{p}^{(1)}
%=
%A J_{1} \left( \frac{\rho {{j}_{0l}}}{a} \right)-
%\sum\nolimits_{m=1}^{\infty } B_m
%J_{1} \left( \frac{\rho {{j}_{0m}}}{a} \right),
%\end{equation}
%%
%where unknowns $ A $ and $ \{ B_m \} $ can be determined after substituting Eq.~\eqref{eq:Psi_p_guess} into Eq.~\eqref{eq:Psi}.
%One obtains:
% 
\begin{equation}
\label{eq:partsol1}
\Psi_p^{ ( 1 ) }( \rho, \alpha )
=
\frac{ i M^{(i)} }{ 2 \pi \sigma_l }
\frac{ J_1 \left( \rho \sigma_l \right) }{ k_{zl} + \alpha } 
-
\sum\limits_{m=1}^{\infty }
\frac{ i M_m }{ 2 \pi \sigma_m }
\frac{ J_1 \left( \rho \sigma_m \right) }{ k_{zm} - \alpha },
\end{equation}
\begin{equation}
\begin{aligned}
\label{eq:partsol2}
&\Psi _{p}^{(2)}( \rho, \alpha )
=
\frac{ i M^{ ( i ) } }{ 2 \pi }
\frac{ \frac{ k_{zl} }{ \varepsilon } - \alpha }{ \kappa^2 - s_l^2 } 
\frac{ J_1 ( b \sigma_l ) }{ \sigma_l \psi_0( s_l ) } \times \\
&\times
\left[ J_1( \rho s_l ) Y_0( a s_l ) - Y_1( \rho s_l ) J_0( a s_l ) \right] - \\
&-
\sum\limits_{m=1}^{\infty }
\frac{ i M_m }{ 2 \pi }
\frac{ k_{ z m } - \alpha }{ \kappa^2 - s_m^2 }
\frac{ J_1 ( b \sigma_m ) }{ \sigma_m \psi_0( s_m ) } \times \\
&\times
\left[ J_1( \rho s_m ) Y_0( a s_m ) - Y_1( \rho s_m ) J_0( a s_m ) \right].
\end{aligned}
\end{equation}
From Eq.~\eqref{eq:Psi2Phi} one obtains:
\begin{align}
\label{eq:gen_sol_Phi_1}
\Phi_+^{ ( 1 ) }( \rho, \alpha ) &= i k_0^{ -1 } C^{(1)} \kappa J_0 ( \rho \kappa ) + \Phi_p^{ ( 1 ) }( \rho, \alpha ), \\
\label{eq:gen_sol_Phi_2}
\Phi_+^{ ( 2 ) }( \rho, \alpha ) &= i k_0^{-1} C^{ ( 2 ) }_1 \kappa J_0 ( \rho \kappa ) + \nonumber \\
&+ i k_0^{ -1 } C^{ ( 2 ) }_2 \kappa J_0( \rho \kappa ) + \Phi_p^{ ( 2 ) }( \rho, \alpha ), \\
\label{eq:gen_sol_Phi_3}
\Phi_-^{ ( 3 ) }( \rho, \alpha ) &+ \Phi_+^{ ( 3 ) }( \rho, \alpha ) = i k_0^{ -1 } C^{ ( 3 ) } \kappa H_1^{ ( 1 ) }( \rho \kappa ),
\end{align}
where
\begin{equation}
\label{eq:partsol_Phi1}
\Phi_p^{ ( 1 ) }( \rho, \alpha )
=
\frac{ - M^{(i)} }{ 2 \pi k_0 }
\frac{ J_0 ( \rho \sigma_l ) }{ k_{ z l } + \alpha } 
+
\sum\limits_{m=1}^{\infty }
\frac{ M_m }{ 2 \pi k_0 }
\frac{ J_0 \left( \rho \sigma_m \right) }{ k_{zm} - \alpha },
\end{equation}
\begin{equation}
\begin{aligned}
\label{eq:partsol_Phi2}
&\Phi_p^{ ( 2 ) }( \rho, \alpha )
=
\frac{ - M^{ ( i ) } }{ 2 \pi k_0 }
\frac{ \frac{ k_{zl} }{ \varepsilon } - \alpha }{ \kappa^2 - s_l^2 } 
\frac{ s_l J_1 ( b \sigma_l ) }{ \sigma_l \psi_0( s_l ) } \times \\
&\times
\left[ J_0( \rho s_l ) Y_0( a s_l ) - Y_0( \rho s_l ) J_0( a s_l ) \right] - \\
&+
\sum\limits_{m=1}^{\infty }
\frac{ M_m }{ 2 \pi k_0 }
\frac{ k_{ z m } - \alpha }{ \kappa^2 - s_m^2 }
\frac{ s_m J_1 ( b \sigma_m ) }{ \sigma_m \psi_0( s_m ) } \times \\
&\times
\left[ J_0( \rho s_m ) Y_0( a s_m ) - Y_0( \rho s_m ) J_0( a s_m ) \right],
\end{aligned}
\end{equation}
and
$ \Phi_p^{ ( 2 ) }( a, \alpha ) = 0 $. 

Boundary condition $ E_{\omega z} = 0 $ for $ \rho = a $, $ z < 0 $ results in $ \Phi_-^{(3)}( a, \alpha ) = 0 $, and we obtain from~\eqref{eq:gen_sol_Phi_3}:
\begin{equation}
\label{eq:C3}
C^{ ( 3 ) }
= \frac{ k_0 \Phi_+^{ ( 3 ) }( a, \alpha ) }
{ i \kappa H_1^{ ( 1 ) }( a \kappa ) },  
\end{equation}
therefore from Eq.~\eqref{eq:gen_sol_3}
\begin{equation}
\label{eq:WHFore}
\Psi_+^{ ( 3 ) }( a, \alpha ) + \Psi_-^{ ( 3 ) }( a, \alpha ) 
= 
\frac{ k_0 \Phi_+^{(3)}(a,\alpha) H_1^{ ( 1 ) }( a \kappa ) }
{ i \kappa H_1^{ ( 1 ) }( a \kappa ) }.
\end{equation}
To obtain Wiener-Hopf-Fock equation one should express the term $ \Psi_+^{ ( 3 ) }( a, \alpha ) $ in Eq.~\eqref{eq:WHFore} through the $ \Phi_+^{(3)}(a,\alpha) $. This can be done as follows. 

First, we use continuity conditions for $ \rho = a $, $ z > 0 $: 
$ H_{ \omega \varphi }^{(3)}( a, z ) = H_{ \omega \varphi }^{(2)}( a, z ) $ 
and 
$ E_{ \omega z }^{(3)}( a, z ) = E_{ \omega z }^{(2)}( a, z ) $ 
therefore  
$ \Psi_+^{(3)}( a, \alpha ) = \Psi_+^{(2)}( a, \alpha ) $
and
$ \Phi_+^{(3)}( a, \alpha ) = \Phi_+^{(2)}( a, \alpha ) $.
Using Eqs.~\eqref{eq:gen_sol_2}, \eqref{eq:gen_sol_Phi_2} and excluding the constant $ C_1^{(2)}$ we have:
\begin{equation}
\label{eq:BCa}
\begin{aligned}
&C_2^{ ( 2 ) } 
=
\left[
\frac{ i \kappa }{ k_0 } 
J_0( a \kappa )
\left(
\Psi_p^{ ( 2 ) }( a, \alpha ) - \Psi_+^{ ( 3 ) }( a, \alpha )
\right) - \right. \\
&-
\left.
J_1( a \kappa )
\left(
\Phi_p^{ ( 2 ) }( a, \alpha ) - \Phi_+^{ ( 3 ) }( a, \alpha ) 
\right)
\vphantom{ \frac{ i \kappa }{ k_0 } }
\right]
\frac{ \pi a k_0 }{ 2 i },
\end{aligned}
\end{equation}
where
$ \Phi_p^{ ( 2 ) }( a, \alpha ) = 0 $, see Eq.~\eqref{eq:partsol_Phi2}.

Second, we use continuity conditions for $ \rho = b $, $ z > 0 $: 
$ H_{ \omega \varphi }^{(2)}( b, z ) = H_{ \omega \varphi }^{(1)}( b, z ) $ 
and 
$ E_{ \omega z }^{(2)}( b, z ) = E_{ \omega z }^{(1)}( b, z ) $ 
therefore  
$ \Psi_+^{(2)}( b, \alpha ) = \Psi_+^{(1)}( b, \alpha ) $
and
$ \Phi_+^{(2)}( b, \alpha ) = \Phi_+^{(1)}( b, \alpha ) $. 
Using Eqs.~\eqref{eq:gen_sol_1}, \eqref{eq:gen_sol_2}, \eqref{eq:gen_sol_Phi_1}, \eqref{eq:gen_sol_Phi_2} and excluding the constants  
$ C^{(1)} $ and $ C_1^{(2)} $ we have:
\begin{equation}
\label{eq:BCb}
\begin{aligned}
&C_2^{(2)}
=
\left[
\frac{ i \kappa }{ k_0 }
J_0( b \kappa )
\left(
\Psi_p^{ ( 2 ) }( b, \alpha ) - \Psi_p^{ ( 1 ) }( b, \alpha )
\right) - \right. \\
&-
\left.
J_1( b \kappa )
\left(
\Phi_p^{ ( 2 ) }( b, \alpha ) - \Phi_p^{ ( 1 ) }( b, \alpha )
\right)
\vphantom{ \frac{ i \kappa }{ k_0 } }
\right]
\frac{ \pi b k_0 }{ 2 i }.
\end{aligned}
\end{equation}
Combining Eqs.~\eqref{eq:BCa} and \eqref{eq:BCb} we obtain the required relation:
\begin{equation}
\label{eq:PsiPlus3}
\begin{aligned}
\Psi_+^{ ( 3 ) }( a, \alpha )
&=
\frac{ k_0 \Phi_+^{ ( 3 ) }( a, \alpha ) J_1( a \kappa ) }
{ i \kappa J_0( a \kappa ) }
+
\Psi_p^{ ( 2 ) }( a, \alpha ) + \\
+ \frac{ b }{ a }
&\left[
\frac{ k_0 J_1( b \kappa ) }{ i \kappa J_0( a \kappa ) } 
\left(
\Phi_p^{ ( 2 ) }( b, \alpha ) - \Phi_p^{ ( 1 ) }( b, \alpha )
\right) - \right. \\
&-\left.
%\vphantom{ \frac{ k_0 J_1( b \kappa ) }{ i \kappa J_0( a \kappa ) } }
\frac{ J_0( b \kappa ) }{ J_0( a \kappa ) }
\left(
\Psi_p^{ ( 2 ) }( b, \alpha ) - \Psi_p^{ ( 1 ) }( b, \alpha )  
\right)
\right].
\end{aligned}
\end{equation}

The following important note should be made here.
It can be checked that the right-hand side of Eq.~\eqref{eq:PsiPlus3} is free from pole singularities for $ \alpha = \pm k_{ z m } $ and for 
$ \alpha $ 
satisfying the equation 
$ \kappa^2(\alpha) = s_m^2 $, 
$ m = 1, 2, \ldots$.
However, the right-hand side of Eq.~\eqref{eq:PsiPlus3} formally possesses pole singularity for 
$ \alpha = \alpha_m $ 
so that 
$ a \kappa( \alpha_m ) = j_{ 0 m } $, 
where 
$ j_{ 0 m } $ 
is the $m$-th zero of Bessel function $ J_0 $, therefore 
$ \alpha_m = \sqrt{ k_0^2 - j_{ 0 m }^2 a^{-2} } $,
$ \mathrm{Im} \alpha_m >0 $ ($ \alpha_m $ are longitudinal wavenumbers of vacuum waveguide of radius $ a $).
%
%Equations~\eqref{eq:Phi} and~\eqref{eq:ErhoEz} result in
%%
%\begin{equation}
%\Phi _{\pm}^{(1,2)} = \frac{ \Psi _{\pm}^{(1,2)} }{\rho}+\frac{ \partial \Psi _{\pm}^{(1,2)} }{\partial \rho},
%\end{equation}
%%
%in particular, one obtains
%%
%\begin{align}
%\Phi _{+}^{(1)}(a, \alpha) &= C_1 \kappa J_0( a \kappa), \\ \nonumber
%\Phi _{+}^{(2)}(a, \alpha) &= C_2 \kappa H_0^{(1)}( a \kappa) \nonumber
%\end{align}
%($\Phi _{-}^{(2)}(a, \alpha) = 0 $ because $ E_{ \omega z } = 0 $ for $ \rho =a $, $ z<0$), therefore
%%
%\begin{equation}
%\label{eq:C1C2}
%\begin{aligned}
%C_1&=\Phi _{+}^{(1)}( a, \alpha) \kappa^{-1} J_0^{-1}( a\kappa ), \\
%C_2&=\Phi _{+}^{(2)}( a, \alpha) \kappa^{-1} \left( H_0^{(1)}( a\kappa ) \right)^{-1}.
%\end{aligned}
%\end{equation}
%%
%Note that
%\begin{equation}
%\label{eq:cont}
%\Psi _{+}^{(1)}(a, \alpha) = \Psi _{+}^{(2)}(a, \alpha),
%\;
%\Phi _{+}^{(1)}(a, \alpha) = \Phi _{+}^{(2)}(a, \alpha)
%\end{equation}
%%
%due to continuity of
%$ E_{ \omega z } $ and $ H_{ \omega \varphi } $ for
%$ \rho = a $,
%$ z > 0 $.
%Using $C_1$ we get from~\eqref{eq:gen_sol_1}, \eqref{eq:partsol} and \eqref{eq:cont}:
%%
%\begin{equation}
%\begin{aligned}
%&  \Psi _{+}^{(1)}(a ,\alpha )=
%\frac{ J_1( a\kappa ) \Phi _{+}^{(2)}( a, \alpha) }{ \kappa J_0 ( a\kappa )} + \frac{i}{2\pi }  \\
%& \times
%\left[
%M^{(i)} \frac{ \frac{ k_{zl} }{\varepsilon} - \alpha }{ \alpha_l^2 -\alpha^2 } J_1 ( j_{ 0 l } ) -
%\sum\limits_{m=1}^{\infty }M_m
%\frac{ \frac{ k_{zm} }{\varepsilon} + \alpha }{ \alpha_m^2 -\alpha^2 } J_1 ( j_{ 0 m } )
% \right].
%\end{aligned}
%\end{equation}
%
However, the function which is determined by Eq.~\eqref{eq:PsiPlus3} should be regular in the area $ \mathrm{Im}\alpha>-\delta $.
Therefore, this pole singularity at the right-hand side should be eliminated and we obtain the following requirement:
\begin{align}
\label{excludepoles}
& \Phi_+^{ ( 3 ) }( a,  \alpha_p )
\frac{ J_1( j_{ 0 p } ) }{ j_{ 0 p } / a }
+
\frac{ b }{ a }
\left\{
\frac{ M^{ ( i ) } }{ 2 \pi k_0 }
\left[
\left( \frac{ k_{ z l } }{ \varepsilon } - \alpha_p \right)
\eta_l( \alpha_p ) - \right. \right. \nonumber \\
&-
\left.
\frac{ \zeta_l( \alpha_p ) }
{ k_{ z l } + \alpha_p }
\right]
-
\sum\limits_{ m = 1 }^{ \infty }
\frac{ M_m }{ 2 \pi k_0 }
\left[
\left( \frac{ k_{ z m } }{ \varepsilon } + \alpha_p \right)
\eta_m( \alpha_p ) - \right. \nonumber \\
&-
\left.
\left.
\frac{ \zeta_m( \alpha_p ) }{ k_{ z m } - \alpha_p }
\right]
\right\}
= 0,
\end{align}
where 
\begin{align}
&\eta_m( \alpha )
=
\frac{ J_1 ( b \sigma_m ) }{ \kappa^2 - s_m^2 } 
\frac{ 1 }{ \sigma_m \psi_0( s_m ) } \times \\
\times
&\left[ 
Y_0( a s_m )
\left(
s_m J_0( b s_m ) \frac{ J_1( b \kappa ) }{ \kappa } - J_0( b \kappa ) J_1( b s_m ) 
\right)
\right. - \label{eq:etaM} \nonumber \\
&-
Y_0( b s_m )
\left(
s_m J_0( a s_m ) \frac{ J_1( b \kappa ) }{ \kappa } - J_0( a \kappa ) J_1( b s_m ) 
\right) + \nonumber \\
&+
\left.
Y_1( b s_m )
\left(
\vphantom{ \frac{ J_1( b \kappa ) }{ \kappa } }
J_0( b \kappa ) J_0( a s_m ) - J_0( b s_m ) J_0( a \kappa ) 
\right)
\right], \nonumber
\end{align}
\begin{equation}
\label{eq:zetaM}
\zeta_m( \alpha )
=
J_0 ( b \sigma_m ) \frac{ J_1 ( b \kappa ) }{ \kappa } 
- 
J_0( b \kappa ) \frac{ J_1( b \sigma_m ) }{ \sigma_m }. 
\end{equation}

Substituting Eq.~\eqref{eq:PsiPlus3} into Eq.~\eqref{eq:WHFore} and combining the terms proportional to $ \Phi_+^{ ( 3 ) }( a,  \alpha ) $ we obtain the following Wiener-Hopf-Fock equation:
\begin{equation}
\label{WHF1}
\frac{ 2 k_0 \Phi_{+}^{ ( 3 ) }( a, \alpha ) }
{ \kappa G(\alpha ) }
+\Psi_-^{ ( 3 ) } ( a, \alpha ) 
+
\frac{ b }{ a }
\frac{ i }{ 2 \pi }
\frac{ \Pi( \alpha ) }{ J_0( a \kappa ) } = 0,
\end{equation}
where
\begin{equation}
\label{eq:G}
G(\alpha )=\pi a\kappa J_0( a \kappa ) H_0^{ ( 1 ) } ( a \kappa ),
\end{equation}
\begin{align}
&\Pi( \alpha )
=
M^{ ( i ) } 
\left[
\left( \frac{ k_{ z l } }{ \varepsilon } - \alpha \right)
\eta_l( \alpha_p ) - 
\frac{ \zeta_l( \alpha ) }
{ k_{ z l } + \alpha }
\right]  - \nonumber \\
&-
\sum\limits_{ m = 1 }^{ \infty }
M_m
\left[
\left( \frac{ k_{ z m } }{ \varepsilon } + \alpha \right)
\eta_m( \alpha ) -
\frac{ \zeta_m( \alpha ) }{ k_{ z m } - \alpha }
\right]. \label{eq:Pi}
\end{align}

Since $G(\alpha)$ and $ \kappa(\alpha) $ are holomorphic and free of zeros and poles in the strip 
$ -\delta < \operatorname{Im}\alpha < +\delta$
one can perform a factorization,
$ \kappa = \kappa_+ \kappa_- $,
where
$ \kappa_{\pm} = \sqrt{ k_0 \pm \alpha } $
and
$ G(\alpha ) = G_+(\alpha ) G_-(\alpha ) $
(standart integral formulas from~\cite{Mittrab} can be used).
Then, Eq.~\eqref{WHF1} should be multiplied by $\kappa_+ G_+$ and consequent decomposition of the function
\begin{equation}
\label{eq:S}
S( \alpha ) = 
\frac{ b }{ a }
\frac{ i }{ 2 \pi }
\Pi( \alpha )
\frac{ \kappa_-( \alpha ) G_-( \alpha ) }{ J_0( a \kappa ) }
\end{equation}
into a sum of ``+'' and ``--'' summands should be performed:
\begin{equation}
\label{eq:Splus}
S_+( \alpha ) = 
\frac{ -i b }{ 2 \pi a }
\sum\limits_{ q = 1 }^{ \infty }
\Pi( -\alpha_q )
\frac{ \kappa_+( \alpha_q ) G_+( \alpha_q ) j_{ 0 q } }{ a^2 \alpha_q J_1( j_{ 0 q } ) ( \alpha + \alpha_q ) },
\end{equation}
$ S_-( \alpha ) = S( \alpha ) - S_+( \alpha ) $.
Then the following equation arises: 
\begin{equation}
\label{WHF2}
\begin{aligned}
&\frac{ 2 k_0 \Phi _{+}^{ ( 3 ) } ( a, \alpha ) }{ \kappa_+( \alpha ) G_+(\alpha ) }
+
S_+( \alpha ) = \\
&=
- S_-( \alpha ) -
\kappa_-( \alpha ) G_-(\alpha ) \Psi _-^{ ( 3 ) } ( a, \alpha ).
\end{aligned}
\end{equation}

Equation~\eqref{WHF2} is solved in a common way~\cite{Mittrab, GVT2021}.
The function on the left hand side of Eq.~\eqref{WHF2} is holomorphic in the area 
$\operatorname{Im}\alpha >-\delta$
while the function on the right hand side is holomorphic in the area 
$\operatorname{Im}\alpha < +\delta$. 
Therefore, due to the analytic continuation theorem~%
\cite{Mittrab},
Eq.~\eqref{WHF2}
determines a function which is holomorphic in the whole complex plane $ \alpha $, this function can be called $ P( \alpha ) $.
From the physical nature of the problem, $ P( \alpha ) $ can be only polynomial which will be clear from the subsequent discussion of Meixner conditions.

To determine $ P( \alpha ) $ one should estimate asymptotic behaviour of all terms in Eq.~\eqref{WHF2} for $ |\alpha | \to \infty $, $-\delta<\mathrm{Im}\alpha < \delta$.
Based on Meixner edge condition~\cite{Mittrab} we have:
\begin{equation}
\label{Meixner}
\begin{aligned}
&\Phi_+^{ ( 3 ) } ( a, \alpha ) \underset{ |\alpha| \to \infty }{ \sim } \alpha^{ -1/2 - \tau }, \; 
\tau = \frac{ 1 }{ \pi } \mathrm{ asin } \frac{ \varepsilon - 1 }{ 2 ( \varepsilon + 1 ) }, \\
& M_m \underset{ m \to \infty }{ \sim } m^{ -1 - \tau }, \; 
\Psi_-^{ ( 3 ) } ( a, \alpha ) \underset{ | \alpha | \to \infty }{ \sim } \alpha^{ -3 / 2 },
\end{aligned}
\end{equation}
therefore all terms in~\eqref{WHF2} decrease in accordance with power law (this consequence illustrates the fact that $ P( \alpha ) $ is a polynomial) and therefore $ P( \alpha ) = 0 $ due to the Liouville's theorem.
Formal solution of the Wiener-Hopf-Fock equation then reads
\begin{equation}
\label{WHFsol}
\begin{aligned}
&\Phi_+^{ ( 3 ) }( a, \alpha ) 
= 
\frac{ - i b }{ 4 \pi k_0 a } \kappa_+ ( \alpha ) G_+( \alpha ) \times \\
&\times
\sum\limits_{ q = 1 }^{ \infty } 
\Pi( -\alpha_q )
\frac{ \kappa_+( \alpha_q ) G_+( \alpha_q ) j_{ 0 q } }{ a^2 \alpha_q J_1( j_{ 0 q } ) ( \alpha + \alpha_q ) }.
\end{aligned}
\end{equation}
It should be noted that $ \Pi( -\alpha_q ) $ in Eq.~\eqref{WHFsol} contains unknown coefficients $ M_m $.
To resolve this, one should substitute~\eqref{WHFsol} into~\eqref{excludepoles}.
After simple but bulky transformations we obtain the following infinite linear system for $\{ M_m \}$:
\begin{equation}
\label{sys}
\sum\nolimits_{ m = 1 }^{ \infty } W_{ p m } M_m = M^{ ( i ) } w_p,
\quad
p = 1, 2, \ldots,
\end{equation}
where
\begin{align}
& W_{ p m } =
\left( \frac{ k_{ z m } }{ \varepsilon } + \alpha_p \right)
\eta_m( \alpha_p ) -
\frac{ \zeta_m( \alpha_p ) }{ k_{ z m } - \alpha_p } + \nonumber \\
&+
\frac{ J_1( j_{ 0 p } ) }{ 2 i j_{ 0 p } / a }
\kappa_+ ( \alpha_p ) G_+( \alpha_p ) \times \nonumber \\
&\times
\sum\limits_{ q = 1 }^{ \infty }
\left[
\left( \frac{ k_{ z m } }{ \varepsilon } - \alpha_q \right)
\eta_m( \alpha_q ) -
\frac{ \zeta_m( \alpha_q ) }{ k_{ z m } + \alpha_q }
\right] \times \nonumber \\
& \times
\frac{ \kappa_+( \alpha_q ) G_+( \alpha_q ) j_{ 0 q } }{ a^2 \alpha_q J_1( j_{ 0 q } ) ( \alpha_p + \alpha_q ) },
\label{W}
\end{align}
\begin{align}
& w_p =
\left( \frac{ k_{ z l } }{ \varepsilon } - \alpha_p \right)
\eta_l( \alpha_p ) -
\frac{ \zeta_l( \alpha_p ) }{ k_{ z l } + \alpha_p } + \nonumber \\
&+
\frac{ J_1( j_{ 0 p } ) }{ 2 i j_{ 0 p } / a }
\kappa_+ ( \alpha_p ) G_+( \alpha_p ) \times \nonumber \\
&\times
\sum\limits_{ q = 1 }^{ \infty }
\left[
\left( \frac{ k_{ z l } }{ \varepsilon } + \alpha_q \right)
\eta_l( \alpha_q ) -
\frac{ \zeta_l( \alpha_q ) }{ k_{ z l } - \alpha_q }
\right] \times \nonumber \\
& \times
\frac{ \kappa_+( \alpha_q ) G_+( \alpha_q ) j_{ 0 q } }{ a^2 \alpha_q J_1( j_{ 0 q } ) ( \alpha_p + \alpha_q ) }.
\label{w}
\end{align}
% 

%For arbitrary finite $ p $ and $ m \to +\infty $ the series in Eqs.~\eqref{W} and \eqref{w} converges because 
%$ W_{ p m } \sim m^{-1} $
%and
%$ M_m \sim m^{ -1 - \tau }$.
%%
%It can be easily shown that for $\varepsilon =1$ this system is analytically solved and the solution coincides with well-known result for open-ended vacuum waveguide~%
%\cite{Weinb}.
%For $\varepsilon \ne 1$ system~\eqref{sys} can be solved numerically using the reducing technique, corresponding procedure is discussed in Sec.~\ref{sec:num}.
The system~\eqref{sys} can be solved numerically using the reducing technique (see, for example~\cite{GVT2021} for details), corresponding results are presented in Sec.~\ref{sec:num}.

\section{EM field derivation}

When the set of coefficients $\{M_m\}$ is determined, the EM field in the domains ``1'', ``2'' and ``3'' can be easily calculated.
We consider in more detail the domain ``3''.
%Combining~\eqref{eq:gen_sol_1} and~\eqref{eq:C1C2} we obtain for the domain ``1'':
%%
%\begin{equation}
%\label{eq:Psi1solved}
%\Psi _{+}^{(1)}(\rho, \alpha)=J_1(\rho \kappa )\frac{\Phi_+^{(2)}(a,\alpha)}{\kappa J_0(a\kappa)}+\Psi _{p}^{(1)}(\rho,\alpha), 
%\end{equation}  
%%
%where a particular solution $\Psi _{p}^{(1)}$ is given by Eq.~\eqref{eq:partsol} while solution for $\Phi_+^{(2)}(a,\alpha)$ is given by Eq.~\eqref{WHFsol}.
Combining~\eqref{eq:gen_sol_3} with~\eqref{eq:C3} we obtain for the domain ``3'': 
\begin{equation}
\label{eq:Psi2solved}
\Psi^{(3)}(\rho,\alpha) = H_{1}^{(1)}(\rho \kappa ) \frac{\Phi_+^{(3)}(a,\alpha)}{\kappa H_0^{(1)}(a\kappa)},
\end{equation}
where solution for $\Phi_+^{(3)}(a,\alpha)$ is given by Eq.~\eqref{WHFsol}.

Field components are calculated via the inverse transform over $\alpha$, in accordance with Eqs.~\eqref{eq:plustrans} and \eqref{eq:minustrans}:
\begin{equation}
\label{eq:Hphi3}
H_{\omega\varphi}^{(3)}(\rho, z) = \int\nolimits_{-\infty}^{+\infty} \Psi^{(3)}(\rho,\alpha) e^{-i\alpha z} \,d\alpha.
\end{equation}

After substitutions, we have:
%%
%\begin{equation}
%\label{eq:Hphi1}
%\begin{aligned}
%&H_{\omega\varphi}^{(1)}(\rho, z) 
%= {-} M^{ ( i ) } J_1( j_{ 0 l } ) \kappa_+( \alpha_l ) G_+( \alpha_l )
%\frac{ \frac{ k_{ z l } }{ \varepsilon } {+} \alpha_l }{ 2 \alpha_l }
%\frac{ I_l^{(1)} }{ 4 \pi } \\
%&{+}
%\sum\limits_{ m {=} 1 }^{\infty} M_m J_1( j_{ 0 m } ) \kappa_+( \alpha_m ) G_+( \alpha_m )
%\frac{ \frac{ k_{ z m } }{ \varepsilon } {-} \alpha_m }{ 2 \alpha_m }
%\frac{ I_m^{(1)} }{ 4 \pi } \\
%&{+}
%\frac{i M^{(i)} J_1(\rho j_{ 0 l } / a )}{2\pi}I_{ p l }^{-} -
%\sum\limits_{m{=}1}^{\infty}\frac{i M_m J_1(\rho j_{ 0 m } / a )}{2\pi}I_{ p m }^{+},
%\end{aligned}
%\end{equation}
%
\begin{equation}
\label{eq:Hphi3}
\begin{aligned}
&H_{\omega\varphi}^{(3)}(\rho, z) 
= 
\frac{ b }{ a }
\sum\limits_{ m = 1 }^{ \infty } 
\Pi( -\alpha_m )
\frac{ \kappa_+( \alpha_m ) G_+( \alpha_m ) j_{ 0 m } }{ a^2 \alpha_m J_1( j_{ 0 m } ) }
\frac{ I_m^{ ( 3 ) } }{ 4 \pi },
\end{aligned}
\end{equation}
where
%
%\begin{equation}
%\label{eq:Im1}
%I_m^{(1)}( \rho, z )
%=
%\pi a \int\limits_{-\infty}^{+\infty} \!\! \frac{ \kappa( \alpha ) J_1( \rho \kappa ) H_0^{(1)}( a \kappa ) }
%{ \kappa_-( \alpha ) G_-( \alpha ) ( \alpha_m + \alpha ) } e^{ -i \alpha z }
%\,d\alpha, 
%\end{equation}
%%
%\begin{equation}
%\label{eq:Ipm}
%I_{ p m }^{ \pm }( z )
%=
%\int\limits_{-\infty}^{+\infty} \!\! 
%\frac{ \frac{ k_{ zm } }{ \varepsilon } \pm \alpha }
%{ \alpha_m^2 - \alpha^2 } 
%e^{ -i \alpha z }
%\,d\alpha, 
%\end{equation}
%
\begin{equation}
\label{eq:Im3}
\begin{aligned}
&I_m^{(3)}( \rho, z )
=
\int\limits_{-\infty}^{+\infty} \!\!
\frac{ \kappa_+( \alpha ) G_+( \alpha ) H_{1}^{(1)}( \rho \kappa ) }
{ \kappa( \alpha ) H_0^{(1)}( a \kappa ) ( \alpha_m + \alpha )}
e^{ -i \alpha z }
\,d\alpha  \\
&=
\pi a \int\limits_{-\infty}^{+\infty} \!\! 
\frac{ \kappa( \alpha ) H_1^{ ( 1 ) } ( \rho \kappa ) J_0( a \kappa ) }
{ \kappa_-( \alpha ) G_-( \alpha ) ( \alpha_m + \alpha ) }
e^{ -i \alpha z }
\,d\alpha.
\end{aligned}
\end{equation}
Hopefully, integral~\eqref{eq:Im3} has been investigated in our previous paper~\cite{GVT2021} (see Eq.~(41)).
For example, it can be easily calculated asymptotically in the far-field area of region ``3'' using saddle point method~\cite{FMb}. 
We consider large distances $ R $ (see Fig.~\ref{fig:geom}) so that $k_0 R \gg 1 $ and angles $ \theta $ satisfying the inequalities $\theta \gg 1/\sqrt{k_0 R}$, $\pi - \theta \gg 1/\sqrt{k_0 R}$. 
%Again, the form of $I_m^{(2)}$ given at the second line of Eq.~\eqref{eq:Im2} should be used.
An asymptotic expansion of $ H_{1}^{(1)}( \rho \kappa )$ for $ \rho |\kappa| \gg 1 $ should be used.
After standard substitutions $ z = R \cos\theta $, $\rho = R \sin\theta$ and introducing a new variable $\xi$ so that $\alpha = k_0 \sin\xi$, $\kappa = k_0\cos\xi$, one obtains the exponential term in the form
\[
\exp \left[ i k_0 R \sin( \theta - \xi ) \right]
\]
which determines the isolated saddle point $\xi_s = \theta - \pi/2$.
Calculating the contribution of $\xi_s$ we obtain for the far-field area:
\begin{equation}
\label{eq:Im3as}
I_m^{ ( 3 ) }( \rho, z )
\approx
\pi a \frac{ e^{i k_0 R } }{ R } 
\frac{ \kappa_-( k_0 \cos \theta ) }{ G_+( k_0 \cos\theta ) } \frac{ 2 J_0( a k_0 \sin \theta ) }{ k_0 \cos \theta {-} \alpha_m },
\end{equation}
which should be substituted to Eq.~\eqref{eq:Hphi3}.

%%%%%%%%%%%%%%%%%%%%%%%%%%%%%%%%%%%%%%%%%%%%%%%
\section{Numerical results\label{sec:num}}
%%%%%%%%%%%%%%%%%%%%%%%%%%%%%%%%%%%%%%%%%%%%%%%
%
%%%%%%%%%%%%%%%%%%%%%%%%%%%%%%
\begin{figure*}[t]
\centering
\includegraphics[width=0.95\textwidth]{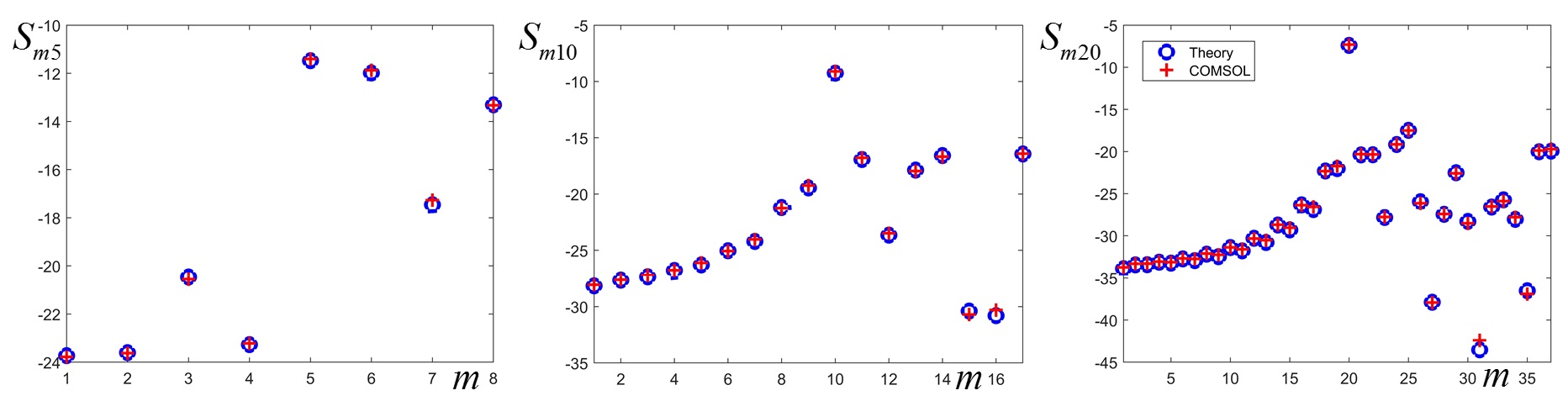}
	\caption{\label{fig:compare} Comparison between $S$-parameters (in dB) obtained via the presented analytical approach and via COMSOL simulations: $ S_{ml} $ corresponds to the Cherenkov frequency  $ f^{\mathrm{CR}}_l $ and incident mode with number $l$. We have 8 propagating modes for $ l = 5 $ ($f^{\mathrm{CR}}_5 = 397$~GHz), 17 for $ l = 10 $ ($f^{\mathrm{CR}}_{10} = 864$~GHz) and 37 for $ l = 20 $ ($f^{\mathrm{CR}}_{20} = 1.81$~THz). Other parameters: $ a = 0.24$~cm, $ b = a/3$, $\varepsilon = 2$.}
\end{figure*}
%%%%%%%%%%%%%%%%%%%%%%%%%%%%%%
%

%%%%%%%%%%%%%%%%%%%%%%%%%%%%%%%
%\begin{figure}[b]
%\centering
%\includegraphics[width=0.95\linewidth]{farfieldCH_rev2}
%\caption{\label{fig:farfieldCH} Far-field distribution of the absolute value of $H_{\omega\varphi}^{(3)}$ in the region ``3'' calculated via Eqs.~\eqref{eq:Hphi3} and \eqref{eq:Im3as}. 
%Observation distance $ R = 500/k_0 $, frequency and other parameters correspond to those in Fig.~\ref{fig:compare}, a number indicated near each curve means the number of the exciting CR mode. 
%$M^{(i)} = 1 $ for each case.
%%The curve $l=20$ is multiplied by the factor $1/2$.
%}
%\end{figure}
%%%%%%%%%%%%%%%%%%%%%%%%%%%%%%%

We solve the system~\eqref{sys} by reducing it to the finite system of $ M_{\max } $ equations, where $ M_{\max } $ was chosen experimentally, around 2-3 times as much as the total number of propagating modes in the waveguide at given frequency.
After that $ M_m $, $ m = 1,2, \ldots M_{\max} $ are immediately calculated, for example, in Matlab.
Such $ M_{\max } $ was typically sufficient since further increase of $ M_{\max } $ resulted in less than 1\% changes in coefficients 
$ M_m $  corresponding to the propagating modes.

For convenient comparison between analytical results and results of numerical simulation, we have calculated powers carrying by incident mode and each reflected propagating mode through the waveguide cross-section.
For this, $z$-component of the Poyting vector averaged over the period $2\pi/\omega$ is calculated (overline means complex conjugation)
\begin{equation}
\label{Szav}
%\begin{aligned}
S^{ \mathrm{ av } }_z = \frac{ c }{ 8 \pi } \Re \left[ E_{ \omega \rho } \overline{ H_{ \omega \varphi } } \right],
%\end{aligned}
\end{equation}
and integrated over the cross-section:
\begin{equation}
\label{powers}
\Sigma = 2 \pi \int\nolimits_0^a S^{ \mathrm{ av } }_z \rho d \rho. 
\end{equation}
Then corresponding $S$-parameters are constructed:
\begin{equation}
\label{spars}
S_{ m l } = \sqrt{ \left. \Sigma_{ m }^{ ( r ) } \right/ \Sigma^{ ( i ) } },
\end{equation}
which also can be expressed in dB, $S^{\mathrm{dB}}_{ m l} = 20 \mathrm{lg}S_{ m l }$.
Note that only propagating modes are essential for the derived $S$-parameters. 
%and the above choice of $M_{\max}$ is valid. 
%However, in other cases, for example, for EM field in the vicinity of the open end where high-order evanescent modes are essential, larger $M_{\max}$ can be required.

Numerical simulations were performed in RF module of COMSOL Multiphysics package. The two dimensional frequency domain solver was utilized. An input end of the waveguide was supported by a series of numerical ports, one separate port for each propagating mode. The port which corresponds to the incident mode was set to be active and option ``active port feedback'' has been disabled. Corresponding eigenmodes were determined numerically, with analytically calculated longitudinal wavenumbers $k_{zm}$ being used as guess values. An open end of the waveguide was surrounded by a semisphere with scattering boundary condition applied. The length of the waveguide and the radius of damping semisphere radius were of the same order, at least several tens of maximum wavelength inside the waveguide.

For calculations of $S$-parameters presented below, the mode frequency was chosen to be equal to the frequency of CR mode $ f^{\mathrm{CR}}_l $ with numbers $l=5$, $l=10$ and $l=20$ produced by a moving charge having with Lorentz factor $\gamma$~\cite{GVT2021}.
%
%\cite{GTAB14}:
%%
%\begin{equation}
%\label{CHfreq}
%\omega^{\mathrm{CR}}_l =
%2 \pi f^{\mathrm{CR}}_l ={c\beta {{j}_{0l}}}/{\left( a\sqrt{\varepsilon {{\beta }^{2}}-1} \right)},
%\end{equation}
%%
%where $\beta =\sqrt{1-{{\gamma }^{-2}}}$.
%To clarify this choice of the frequency let us discuss the relation of the obtained results to the problem of diffraction of a charged particle bunch wakefield at the open-end of the discussed dielectric-loaded waveguide.
%Wakefield is a CR generated inside a wavegiude as an infinite set of discrete frequencies~\eqref{CHfreq}, while each frequency contribution to the total field is usually referred to as a CR mode.
%A CR mode can be presented (after simplification) in the following form~\cite{B62}:
%%
%\begin{equation}
%\label{CHmode}
%H_{ \varphi l }^{\mathrm{CR}}
%=
%\mathrm{Im} \left[  i H_{\varphi 0 l } J_1 \left( \rho \frac{ j_{0l} }{a} \right) e^{ \frac{ i \omega^{\mathrm{CR}}_l z }{ c \beta } } e^{ - i \omega^{\mathrm{CR}}_l t } \right],
%\end{equation}
%%
%where $ c \beta $ is bunch velocity, $H_{\varphi 0 l }$ is some constant.
Since for $ \omega = \omega^{\mathrm{CR}}_l $ we have $ k_{ z l } =  \omega^{\mathrm{CR}}_l  / ( c \beta ) $, an incident mode~\eqref{eq:Hphii} corresponds to the $l$-th CR mode if $ M^{(i)} $ is chosen appropriately.

Figure~\ref{fig:compare} shows comparison between $S$-parameters calculated via presented rigorous analytical approach and obtained from COMSOL simulations.
As one can see, the agreement between results is excellent. This fact proves the presented theory and also shows correctness of COMSOL simulation procedure. One can see that for large enough $l$ ($l=10$ and $l=20$ in Fig.~\ref{fig:compare}) the reflected mode with the number of incident mode dominates (it has the largest $S$-parameter), therefore the overall diffraction process is similar to a single mode reflection. However, for lower $l$ ($l=5$) other modes (especially those with close numbers) can be significant and therefore can alter mentioned ``close to single mode'' regime.

\section{Conclusion}

We have presented an elegant and convenient rigorous analytical approach for calculation of various diffraction processes at the open end (with orthogonal cut) of a circular waveguide with dielectric lining.
The obtained results have been compared to the results of simulations with commercial code COMSOL and an excellent agreement has been observed.
In this paper, we have considered the problem with layered dielectric filling of the waveguide and excitation by single mode of Cherenkov wakefield which is relevant to a series of prospective beam and THz applications mentioned in the Introduction.
However, this powerful approach can be utilized for solving other similar problems.
For example, excitation by a charged particle bunch (in full formulation including both wakefield and Coulomb field) or by an external electromagnetic wave can be incorporated into the solution.

%It is worth noting that computation resources used by the Matlab code based on analytical formulas is much smaller then those occupied by COMSOL.
%Therefore this code can be easily run on ordinary PCs.
%This fact makes the discussed rigorous approach extremely useful for further development of various prospective applications based on electromagnetic interactions of single-cycle THz pulses, wide-band THz wakefields and charged particle bunches with dielectric-lined waveguide structures.
%
%Besides mentioned open-ended waveguides with straight cut, this method can be also useful for investigation of structures with non-orthogonal cut.
%Since in this case the rigorous theory can be marginally applied (solution for moreless similar parallel-plate dielectric-loaded waveguide problem has been reported just recently~\cite{Daniele2019}), development of reliable approximate methods is the most substantial idea (see, for example, \cite{GTAB14}).
%Such reliable methods can be benchmarked and adjusted at simpler structures with orthogonal end cut.

\section{Acknowledgements}

This work is supported by the Russian Science Foundation (grant No. 18-72-10137).

%\bibliographystyle{IEEEtran}
%\bibliography{SNG_Bibliography_Mar2021}
%

\end{document}